\documentclass[preprintnumbers,nofootinbib,superscriptaddress,10.5pt]{revtex4}
\usepackage[dvipdfmx]{graphicx} 
\usepackage{amsmath,amssymb,amsfonts, bm} 

\def\a{\alpha} \def\b{\beta} \def\g{\gamma}  \def\d{\delta}     \def\th{\theta}   \def\l{\lambda}        \def\r{\rho}          

\def\dg{\dagger}  \def\nn{\nonumber}

    \newcommand{\To}{\Rightarrow}

\def\abs#1{\left| #1\right|}

\renewcommand{\Im}{{\rm Im}\,}

\newcommand{\Diag}[3]{ \begin{pmatrix} #1 & 0 & 0 \\ 0 & #2 & 0 \\ 0 & 0 & #3 \\\end{pmatrix}}

\usepackage{color}

\begin{document}
%%%%%%%%%%%%%%%%%%%%%%%%%%%%%%%%%%%%%%%%

\title{\large Direct Evaluation of CP Phase in CKM matrix, \\ General Perturbative Expansion and Relations with  Unitarity Triangles}

\preprint{STUPP-25-283}
%%%%%%%%%%%%%%%%%%%%%%%%%%%%%%%%%%%%%%%%

\author{Masaki J. S. Yang}
\email{mjsyang@mail.saitama-u.ac.jp}
\affiliation{Department of Physics, Saitama University, 
Shimo-okubo, Sakura-ku, Saitama, 338-8570, Japan}
\affiliation{Department of Physics, Graduate School of Engineering Science,
Yokohama National University, Yokohama, 240-8501, Japan}

%%%%%%%%%%%%%%%%%%%%%%%%%%%%%%%%%%%%%%%%

%\date{\today}

%%%%%%%%%%%%%%%%%%%%%%%%%%%%%
\begin{abstract} %%%%%%%%%%%%%%%%%%%%%
%%%%%%%%%%%%%%%%%%%%%%%%%%%%%

In this letter,  using a rephasing invariant formula 
$\delta =  \arg [ { V_{ud} V_{us} V_{c b} V_{tb} / V_{ub} \det V_{\rm CKM} }]$, 
we evaluate the CP phase $\delta$ in the CKM matrix $V_{\rm CKM}$ perturbatively for small quark mixing angles $s_{ij}^{u,d}$ with associated phases $\rho_{ij}^{u,d}$.
 Consequently, we derived a relation 
$\delta \simeq \arg [\Delta s_{12} \Delta s_{23} / ( \Delta s_{13} - s^u_{12} e^{-i \rho^u_{12}}  \Delta s_{23} )]$
with $\Delta s_{ij} \equiv s^d_{ij} e^{-i \rho^d_{ij}} - s^u_{ij} e^{-i \rho^u_{ij}}$, 
which represents the analytic behavior of the CKM phase. 
The uncertainty in the relation is of order $O(\lambda^{2}) \sim 4\%$, comparable to the current experimental precision. 
Experimental data suggest that the hypothesis of some CP phases being maximal.
We also discussed relationships between the phase $\delta$ and unitarity triangles. 
As a result, several relations between the angles $\alpha, \beta, \gamma$ and $\delta$ are identified through other invariants $V_{il} V_{jm} V_{kn} / \det V_{\rm CKM}$. 

%%%%%%%%%%%%%%%%%%%%%%%%%%%%%
\end{abstract} %%%%%%%%%%%%%%%%%%%%%%
%%%%%%%%%%%%%%%%%%%%%%%%%%%%%

\maketitle

%%%%%%%%%%%%%%%
\section{Introduction}
%%%%%%%%%%%%%%%

Understanding the origin of CP violation is crucial for explaining the baryon asymmetry in the universe. 
In the Standard Model, the CP violation in the quark sector is described by a single phase contained in the Cabibbo--Kobayashi--Maskawa (CKM) matrix \cite{Cabibbo:1963yz, Kobayashi:1973fv}. 
Although the mixing matrix depends on the choice of parameterization, 
physical observables must be formulated in terms of rephasing invariants 
\cite{Wu:1985ea, Bjorken:1987tr, Nieves:1987pp, Branco:1999fs, Kuo:2005pf, Jenkins:2007ip, Chiu:2012uc,Chiu:2015ega}. 

Traditionally, the Jarlskog invariant $J$ \cite{Jarlskog:1985ht} has played a central role in the description of CP violation. 
In the discussion of weak basis invariants (WBI) \cite{Bernabeu:1986fc, Gronau:1986xb, Botella:1994cs}, it has been shown that the lowest-dimensional CP-odd WBI is proportional to $J$. 
However, since the invariant $J$ is a small quantity of order $O(10^{-5})$, 
it is highly sensitive to various approximations. 
Capturing analytic behavior of the CP violation---including error estimation---has been a technically challenging task due to the computational precision.
Although Ref.~\cite{Hall:1993ni} presented a general perturbative treatment of the CKM matrix and its CP phase, such a general treatment has not been pursued in later literature.

In recent years, a rephasing invariant involving the determinant of the mixing matrix
 has been proposed for a direct evaluation of the CP phase~\cite{Yang:2025hex}. 
 It complements conventional approaches in both of analytic structure and error estimation.
In this letter, employing the rephasing invariant formula $\delta =  \arg [ { V_{ud} V_{us} V_{c b} V_{tb} / V_{ub} \det V_{\rm CKM} }]$, 
we derive a general perturbative expression for the CP phase in the CKM matrix, 
and its phenomenological consequences. 
This allows for precise analyses that were difficult to achieve with the Jarlskog invariant, which is a cubicly suppressed quantity by the three mixing angles. 
Furthermore, the insights from the quark sector are applicable to the analysis of the leptonic CP phase $\delta_l$. 
With forthcoming experiments, DUNE \cite{DUNE:2020jqi} and T2HK \cite{Hyper-KamiokandeProto-:2015xww}, such analyses are anticipated to shed light on underlying grand unified  relations.

%This paper is organized as follows. 
%The next section gives a review of DRS in the quark sector. 
%
%In Sec.~3, we discuss a field theoretical realization of DRS by the type-I seesaw mechanism. 
%
%In Sec.~4, a systematic analysis of the CKM matrix with DRS is performed. 
%The final section is devoted to a summary. 

%%%%%%%%%%%%%%%%%%%&&&&&&&&&
\section{A Rephasing Invariant formula for CP Phase of  CKM Matrix and Its Perturbative Expansion}
%%%%%%%%%%%%%%%%%%%&&&&&&&&&

We begin by presenting a method to directly extract the CP phase $\delta$ 
 from the CKM matrix $V_{\rm CKM}$ defined in a general phase basis.
To convert a given $V_{\rm CKM}$ into the PDG standard parametrization $V_{\rm CKM}^{0}$, 
unphysical phases are removed by redefinition of phases as
\begin{align}
V_{\rm CKM}^{0} = \Psi_{L}^{\dagger} V_{\rm CKM} \Psi_{R} \, .
\end{align}
Here, $\Psi_{L,R} = {\rm diag}(e^{i \gamma_{(L,R)1}},\, e^{i \gamma_{(L,R)2}},\, e^{i \gamma_{(L,R)3}})$ 
are diagonal phase matrices, and each $\gamma_{(L,R)i}$ represents phases to be solved.  
The inverse transformation from the PDG parametrization to the original basis $V_{\rm CKM} = \Psi_{L} V_{\rm CKM}^{0} \Psi_{R} ^{\dagger}$ is  written by
\begin{align}
\begin{pmatrix}
V_{ud} & V_{us} & V_{ub}  \\
V_{cd} & V_{cs} & V_{cb}  \\
V_{td} & V_{ts} & V_{tb}  \\
\end{pmatrix}
= 
\Diag{e^{i \g_{L1}}}{e^{i \g_{L2}}}{e^{i \g_{L3}}}
\begin{pmatrix}
|V_{ud}| & |V_{us}| & |V_{ub}| e^{-i\d}  \\
V_{cd}^{0} & V_{cs}^{0} & |V_{cb}|  \\
V_{td}^{0} & V_{ts}^{0} & |V_{tb}| \\
\end{pmatrix}
\Diag{e^{- i \g_{R1}}}{e^{- i \g_{R2}}}{e^{- i \g_{R3}}} . 
\end{align}
In particular, focusing on $V_{ub} = |V_{ub}| e^{-i (\delta - \gamma_{L1} + \gamma_{R 3})}$, 
\begin{align}
V_{ub}^{*} = |V_{ub}| e^{i(\d - \g_{L1} + \g_{R3})} \, , ~~ 
V_{ub}^{*}e^{i(\g_{L1} - \g_{R3})} = |V_{ub}| e^{i \d } \, .
\end{align}
By using $\det V_{\rm CKM} = e^{i(\g_{L1} + \g_{L2} + \g_{L3} - \g_{R1} - \g_{R2} - \g_{R3})}$, 
the remaining four phases are eliminated as
\begin{align}
\arg \left[ {V_{ud} V_{us} V_{cb} V_{tb}  \over \det V_{\rm CKM}} \right]
= \g_{L1} - \g_{R3}  \, , ~~ \To ~~ 
\d = \arg \left[ {V_{ud} V_{us} V_{cb} V_{tb}  \over V_{ub} \det V_{\rm CKM}} \right] \, . 
\end{align}
Including the phase of $\det V_{\rm CKM}$, this formula is explicitly rephasing invariant, and it clearly coincides with the phase $\delta$ in the PDG parametrization. 
Except for this particular combination, the phase $\delta$ cannot be solved explicitly because it remains on the right-hand side. 
The subsequent papers \cite{Yang:2025ftl, Yang:2025vrs} show that 
other similar combinations correspond to CP phases in other parametrizations of the mixing matrix. 

The formula possesses several advantages over the traditional Jarlskog invariant \cite{Jarlskog:1985ht}, as summarized below:
\begin{description}
\item[1. Factorizability:]

The CP phase is decomposed into individual elements $V_{\alpha i}$ and the determinant $\det V_{\rm CKM}$, making the computation more transparent and straightforward. 
Furthermore, we can easily quantify uncertainties and errors of the phase. 
Although the Jarlskog invariant $J$ is highly sensitive to approximations because of its smallness,  
the new invariant is of order $O(1)$ and therefore less affected by perturbative corrections. 

\item[2. Independence of constraints:] 

The invariant $J$ simultaneously constrains the mixing angles and the phases. 
 In this formula, since the phase depends only on arguments of matrix elements, 
 the constraint is independent of those of mixing angles.

\item[3. Completeness of phase information:]
The invariant $J$ does not retain the sign of $\cos\delta$, requiring additional calculations to reconstruct the full experimental results. In contrast, this formula directly preserves the full information on the CP-violating phase $\delta$.
\end{description}

We demonstrate the relation between this formula and the well-known Jarlskog invariant.
By dividing the complex quantity (inside the argument) by its modulus, the phase is reconstructed as 
\begin{align}
e^{ i \d} = \frac{ V_{ud} V_{us} V_{c b} V_{tb} }{ V_{ub} \det V_{\rm CKM} }  
\abs{ \frac{ V_{ub} \det V_{\rm CKM} }{ V_{ud} V_{us} V_{c b} V_{tb} } }
=  \frac{ V_{ud} V_{us} V_{c b} V_{tb} V_{ub}^{*} \det V_{\rm CKM}^{*}}{ |V_{ud} V_{us} V_{c b} V_{tb} V_{ub}| } 
 \, . 
\end{align}
Here, we used the identity $ \det V_{\rm CKM}  \det V_{\rm CKM}^{*} = | \det V_{\rm CKM}|^{2} = 1$. 
Since our goal is to obtain the invariant $V_{ud} V_{tb} V_{ub}^* V_{td}^*$,
an alternative element of the inverse matrix $V_{us} \det V_{\rm CKM}^{*} = V_{cb}^* V_{td}^* - V_{cd}^* V_{tb}^*$ yields 
\begin{align}
e^{ i \d} = \frac{ V_{ud} V_{c b} V_{tb} V_{ub}^{*} (V_{cb}^* V_{td}^* - V_{cd}^* V_{tb}^*) }{ |V_{ud} V_{us} V_{c b} V_{tb} V_{ub}| } 
= \frac{ |V_{c b}|^{2} V_{ud}  V_{tb} V_{ub}^{*} V_{td}^*
 - |V_{tb}|^{2} V_{ud} V_{c b} V_{ub}^{*}  V_{cd}^*  }{ |V_{ud} V_{us} V_{c b} V_{tb} V_{ub}| }  \, . 
\end{align}
Taking the imaginary parts of both sides, the right-hand side contains the Jarlskog invariant
$J \equiv  \Im [V_{ud}  V_{tb} V_{ub}^{*} V_{td}^*]$. 
Using the orthonormal relation $|V_{cb}|^{2} + |V_{tb}|^{2} = 1 - |V_{ub}|^{2}$,
we obtain 
\begin{align}
\Im e^{i\d} =  \frac{  (1 - |V_{ub}^{2}| ) J}{ |V_{ud} V_{us} V_{c b} V_{tb} V_{ub}| } 
= \sin \d \, . 
\end{align}
The last equality  follows directly from an expression $J = c_{12} s_{12} c_{23} s_{23} c_{13}^{2} s_{13} \sin \delta$ with the observed mixing angles $s_{ij}, c_{ij}$. 
Therefore, this $\sin \delta$ coincides with the value  derived from the Jarlskog invariant.

Next, we perform the perturbative expansion.
The CKM matrix $V_{\rm CKM} \equiv U_{u}^{\dg} U_{d}$ is defined as the misalignment between the diagonalization matrices of the left-handed up-type quarks $U_{u}$ and down-type quarks $U_{d}$. 
By choosing an appropriate basis, elements of both 
$U_{u,d}$ are taken to be of the same order as those of $V_{\rm CKM}$ without loss of generality.
Therefore, we adopt the following approximation.

\begin{description}

\item[\textbf{Approximation:}]
The mixing angles $s_{ij}^{u,d} \equiv \sin \th_{ij}^{u,d} \, , \, c_{ij}^{u,d} \equiv \cos \th_{ij}^{u,d}$ of $U_{u,d}$ satisfy $s_{12}^{u,d} \sim \lambda$, $s_{23}^{u,d} \sim \lambda^{2}$, and $s_{13}^{u,d} \sim \lambda^{3}$, with an expansion parameter $\lambda \simeq 0.2$.  

\item[\textbf{Justification:}]

When the Yukawa matrices $Y_{u,d}$ of quarks possess chiral symmetries for the first and second generations,
$Y_{u,d} = D_{L} Y_{u,d} D_{R}$, all lighter singular values and mixings vanish. 
Here, $D_{L,R} \equiv {\rm diag} (e^{i \phi_{L,R}^{1}}, e^{i \phi_{L,R}^{2}}, 1)$ and $\phi_{L,R}^{1,2}$ are phases. 
Although these chiral symmetries are only approximate in reality, the mixing angles are suppressed by powers of corresponding ratios of singular values $m_{fi} / m_{fj}$.

\end{description}

We now proceed to define the notation of  perturbative expansion. 
The matrices $U_{u,d}$ are generally written as 
$U_{u,d} = \Phi_{u,d}^{L} U_{u,d}^{0} \Phi_{u,d}^{R}$ 
with diagonal phase matrices $\Phi_{u,d}^{L,R}$ and their PDG parametrizations $U_{u,d}^{0}$. 
Due to the freedom of right-handed phase transformations, the unitary matrices are redefined as
$U_{u,d}^{1} = \Phi_{u,d}^{L} U_{u,d}^{0} \Phi_{u,d}^{L \dagger}$.  
Since $U_{u,d}^{1}$ has a unit determinant, it is written as 
\begin{align}
U_{u,d}^{1} = 
\begin{pmatrix}
1 & 0& 0 \\
0& c_{23}^{u,d} & s_{23}^{u,d}  e^{- i \r_{23}^{u,d} } \\
0& - s_{23}^{u,d}  e^{i \r_{23}^{u,d} } & c_{23}^{u,d}  \\
\end{pmatrix} 
\begin{pmatrix}
c_{13}^{u,d}  & 0 & s_{13}^{u,d}  e^{- i \r_{13}^{u,d} } \\
0 & 1 & 0 \\
- s_{13}^{u,d}  e^{i \r_{13}^{u,d} }& 0 & c_{13}^{u,d}  \\
\end{pmatrix} 
\begin{pmatrix}
c_{12}^{u,d}  & s_{12}^{u,d} e^{- i \r_{12}^{u,d} } & 0\\
- s_{12}^{u,d}  e^{i \r_{12}^{u,d} }& c_{12}^{u,d}  &0 \\
0 & 0 & 1 \\
\end{pmatrix}  ,
\end{align}
where $\r_{ij}^{u,d}$ are the associated CP-violating phases corresponding to the mixing angles.
Under this setup,  the leading order of the mixing matrix is approximated as 
\begin{align}
U_{u,d}^{1} 
 \simeq 
\begin{pmatrix}
1 & s_{12}^{u,d} e^{- i \r_{12}^{u,d} } & s_{13}^{u,d}  e^{- i \r_{13}^{u,d} } \\
- s_{12}^{u,d}  e^{i \r_{12}^{u,d} }& 1 & s_{23}^{u,d}  e^{- i \r_{23}^{u,d} } \\
- s_{13}^{u,d}  e^{i \r_{13}^{u,d} } + s_{12}^{u,d} s_{23}^{u,d}   e^{i \r_{12}^{u,d} + i \r_{23}^{u,d} }& - s_{23}^{u,d}  e^{i \r_{23}^{u,d} } & 1 \\
\end{pmatrix}  . 
\end{align}
The next-to-leading order terms in each matrix element are suppressed by at least order $\lambda^{2}$ compared to the leading order.
Since the right-handed phases $\Phi^{R}_{u,d}$ of quarks do not affect the observed CP phase, we will omit  them hereafter.

In this case, the CKM matrix to be analyzed is redefined as
$V_{\rm CKM} \equiv U_{u}^{1 \dagger} U_{d}^{1}$.
Expanding arguments of each matrix element in powers of $\lambda$, we obtain
\begin{align}
\arg V_{ud} & = 0 + O(\l^{2}) \, , ~~~ \arg V_{tb} = 0 + O(\l^{4}) \, ,   \\
%%%%%%%%%%%%%%%%%%%%
\arg V_{us} & = \arg \left[  s^d_{12} e^{-i \rho^d_{12}} - s^u_{12} e^{-i \rho^u_{12}} \right]+ O(\l^{2}) \, ,   ~~~
 %%%%%%%%%%%%%%%%%%%%
\arg V_{cb} = \arg \left[ s^d_{23} e^{-i \rho^d_{23} } - s^u_{23} e^{-i \rho^u_{23}} \right]+ O(\l^{2}) \, ,  \label{VusVcb} \\
 %%%%%%%%%%%%%%%%%%%%
\arg V_{ub} & = \arg \left[ s^d_{13} e^{-i \rho^d_{13}} - s^u_{13} e^{-i \rho^u_{13}}
- s^u_{12} e^{-i \rho^u_{12}} (s^d_{23} e^{-i \rho^d_{23}} - s^u_{23} e^{-i \rho^u_{23}} ) \right]+ O(\l^{2})\, . 
 %%%%%%%%%%%%%%%%%%%%
\end{align}
In the limit $s_{ij}^{u,d} \to 0$, the associated CP phases $\r_{ij}^{u,d}$ simultaneously vanish.
Therefore, the contributions from these CP phases appear at first order in the mixing angles.
Note that observed mixing angles $s_{ij}$ and $c_{ij}$ constraint the absolute values of matrix elements as, 
\begin{align}
|V_{us}| =  s_{12} c_{13} \, , ~~
|V_{ub}| = s_{13} \, , ~~
|V_{cb}| = s_{23} c_{13} \, .
\end{align}

From the rephasing invariant formula, a general perturbative relation for the CP phase of the CKM matrix will be 
\begin{align}
\d &= \arg [V_{us} V_{cb} / V_{ub}] + O(\l^{2}) \, ,  \nn \\
\abs{V_{us} \over V_{ub} / V_{cb} }e^{i \d} & =  \frac{ s^d_{12} e^{-i \rho^d_{12}} - s^u_{12} e^{-i \rho^u_{12}} } 
{ \dfrac{s^d_{13} e^{-i \rho^d_{13}} - s^u_{13} e^{-i \rho^u_{13}} }{ s^d_{23} e^{-i \rho^d_{23}} - s^u_{23} e^{-i \rho^u_{23}} }
 - s^u_{12} e^{-i \rho^u_{12}}  }  + O(\l^{2}) \, .
\end{align}
The errors of this expression are $O(\lambda^{2}) \simeq 4 \%$, 
which is comparable to the current experimental uncertainties.
In particular, the absolute value of the denominator has a fixed value, 
\begin{align}
\abs{ \dfrac{s^d_{13} e^{-i \rho^d_{13}} - s^u_{13} e^{-i \rho^u_{13}} }{ s^d_{23} e^{-i \rho^d_{23}} - s^u_{23} e^{-i \rho^u_{23}} }  - s^u_{12} e^{-i \rho^u_{12}}  } = \abs{V_{ub} \over V_{cb}} = 0.09 \, . 
\end{align}
Thus, depending on the magnitude of $s_{12}^{u}$, one of the terms in the denominator can be neglected.

It is theoretically intriguing to investigate origins of the observed large CP phase. 
In situations where the 1-3 mixing angles $s_{13}^{u,d}$
are sufficiently smaller than $s_{12}^{u} V_{cb}$ by the above chiral symmetries, 
neglect of these terms yields 
\begin{align}
\abs{V_{us} \over V_{ub} / V_{cb} }e^{i \d} & \simeq 
\frac{ s^d_{12} e^{-i \rho^d_{12}} - s^u_{12} e^{-i \rho^u_{12}} }{  - s^u_{12} e^{-i \rho^u_{12}}  }  + O(\l^{2}) \, , \nn \\
\d & \simeq \arg \left[ 1 - { s^d_{12}  \over s^u_{12} } e^{ i ( \rho^u_{12} - \rho^d_{12})} \right] + O(\l^{2}) \, .
\end{align}
The observed physical value of the CKM matrix in the latest UTfit is  \cite{UTfit:2022hsi}
\begin{align}
\sin \th_{12}^{\rm CKM} &= 0.22519 \pm 0.00083 \, ,  ~~~ \sin \th_{23}^{\rm CKM} = 0.04200 \pm 0.00047 \, , \nn \\
\sin \th_{13}^{\rm CKM} &= 0.003714 \pm 0.000092 \, ,  ~~~ \d = 1.137 \pm 0.022 = 65.15 \pm1.3^{\circ}   \, . 
\end{align}
Given this experimental value, the hypothesis that the phase  is maximal $ \rho^d_{12} - \rho^u_{12} = \pi /2 $ appears quite plausible~\cite{Shin:1985cg, Gronau:1985tx, Fritzsch:1985yv, Kang:1985nw,  Lehmann:1995br, Kang:1997uv, Fritzsch:1999im, Antusch:2009hq, Xing:2003yj, Yang:2020qsa, Yang:2020goc, Yang:2021smh}.

By combining these results with the previous work, 
predictions of grand unified relations become much more straightforward. 
The general perturbative evaluation of the leptonic CP phase $\delta_{l}$ is approximately \cite{Yang:2025hex}
\begin{align}
\delta_{l} \simeq 
\delta _{\nu } + 2 s^e_{23}  \sin \rho _{23}^{e} 
+  4.66 s^e_{12} \sin (\delta _{\nu } - \rho _{12}^{e} )  
+ 4.66 s^e_{13}  \sin (\delta _{\nu } - \rho _{13}^{e} )  
 \, ,
\end{align}
where $\delta_{\nu}$ is the CP phase in the PDG parametrization of the neutrino diagonalization matrix $U_{\nu}$.
The mixing angles $s^{e}_{ij}$ and CP phases $\rho^{e}_{ij}$  for the charged leptons 
are defined in the same way as for the quarks. 
If a large 1-2 mixing $s^e_{12} \sim 0.1$ and the associated phase $\rho_{12}^{e} \sim \pi/2$ exist as in the quark sector, they can contribute approximately $0.5 \sim 30^{\circ}$ to $\delta_{l}$. 
Therefore, without accidental cancellations, 
the CP phase $\delta_{l}$ is expected to be discovered within roughly ten years of operations
 of DUNE \cite{DUNE:2020jqi} and T2HK \cite{Hyper-KamiokandeProto-:2015xww}.
Non-perturbative treatments beyond perturbation theory for $s_{12}^{f}$ and $s_{23}^{f}$ are also available in previous papers~\cite{Yang:2024ulq, Yang:2025yst}.

%%%%%%%%%%%%%%
\subsection*{Relation between the phase and unitarity triangles}
%%%%%%%%%%%%%%

It is of particular interest to discuss  relations between the results  and unitarity triangles 
\cite{Hocker:2006xb, Frampton:2010ii, Dueck:2010fa, Li:2010ae}. 
From the alternative $b$-$s$ unitarity triangle, given by 
$V_{ub}^{} V_{us}^{*} + V_{cb}^{} V_{cs}^{*} +V_{tb}^{}  V_{ts}^{*} = 0$,
its three angles are rephasing invariant quantities \cite{Wu:1994di,Harrison:2006bj,Harrison:2009bz}, 
\begin{align}
\a' = \arg \left [ - { V_{tb}^{} V_{ts}^{*} \over V_{ub}^{} V_{us}^{*} } \right ] = 64.09^{\circ}, ~~
\b' = \arg \left [ - { V_{cb}^{} V_{cs}^{*} \over V_{tb}^{} V_{ts}^{*} }  \right ] = 1.05^{\circ}, ~~
\g' = \arg \left [ - { V_{ub}^{} V_{us}^{*} \over V_{cb}^{} V_{cs}^{*} }  \right ] = 114.86^{\circ}. 
\end{align}
If the phase of $V_{cs}$ can be neglected, the phase $\d$ is related to one of angles of the unitarity triangle.
\begin{align}
\pi -  \g'  =  \arg \left [ { V_{us}^{} V_{cb}^{} \over  V_{ub}^{} V_{cs}^{}} \right ] 
= 65.15^{\circ} \simeq  \d  \, .  %\arg \left[ {V_{us} V_{cb} \over  V_{ub}} \right] - \arg V_{cs}^{} \, , ,,,
\end{align}
Indeed, relations between the phase and the angles are
\begin{align}
\d - \a'  =  \arg \left[ \frac{ V_{ud} V_{us} V_{c b} V_{tb} }{ V_{ub} \det V_{\rm CKM} } \right]
- \arg \left [ - { V_{tb}^{} V_{ts}^{*} \over V_{ub}^{} V_{us}^{*} } \right ]
=  \arg \left[- \frac{V_{ud} V_{c b} V_{ts} }{  \det V_{\rm CKM} } \right] 
= 1.05^{\circ} \, ,  \nn \\
\d + \g' - \pi =  \arg \left[ \frac{ V_{ud} V_{us} V_{c b} V_{tb} }{ V_{ub} \det V_{\rm CKM} } \right]
+ \arg \left [ { V_{ub} V_{cs} \over V_{cb} V_{us}} \right ]   
=  \arg \left[ \frac{ V_{ud} V_{cs} V_{tb} }{  \det V_{\rm CKM} } \right] 
= - 0.0019^{\circ}
\, , 
\end{align}
which define other rephasing invariants.
Moreover, the angles of the standard unitarity triangle are
\begin{align}
\a = \arg \left [ - { V_{td}^{} V_{tb}^{*} \over V_{ud}^{} V_{ub}^{*} } \right ] = 92.40^{\circ}, ~~
\b = \arg \left [ - { V_{cd}^{} V_{cb}^{*} \over V_{td}^{} V_{tb}^{*} }  \right ] = 22.49^{\circ}, ~~
\g = \arg \left [ - { V_{ud}^{} V_{ub}^{*} \over V_{cd}^{} V_{cb}^{*} }  \right ] = 65.11 ^{\circ},
\end{align}
and exact relations with these angles are found to be 
\begin{align}
\d + \a + \pi & =  \arg \left[ \frac{ V_{ud} V_{us} V_{c b} V_{tb} }{ V_{ub} \det V_{\rm CKM} } \right] + \arg \left [ { V_{td}^{} V_{tb}^{*} \over V_{ud}^{} V_{ub}^{*} } \right ]
=  \arg \left[ \frac{  V_{us} V_{c b} V_{td} }{  \det V_{\rm CKM} } \right] = - 22.45^{\circ} \fallingdotseq  - \b\, ,  \nn \\
\d - \g - \pi & = \arg \left[ \frac{ V_{ud} V_{us} V_{c b} V_{tb} }{ V_{ub} \det V_{\rm CKM} } \right] +
\arg \left [ { V_{ud}^{} V_{ub}^{*} \over V_{cd}^{} V_{cb}^{*} }  \right ]
= \arg \left[ \frac{  V_{us} V_{cd} V_{tb} }{ \det V_{\rm CKM} } \right]
= -179.96 \fallingdotseq - \pi . 
\end{align}
These two expressions are not independent; from $\arg 2\pi = 0$, we obtain 
\begin{align}
\epsilon \equiv \d + \a + \b + \pi = \d - \g = \arg \left[ - \frac{  V_{us} V_{cd} V_{tb} }{ \det V_{\rm CKM} } \right]
 = 0.035^{\circ} \, . 
\end{align}
Since the perturbative expansions of $V_{us}$ and $V_{cd}$ agree up to $\mathcal{O}(\lambda^3)$ in Eq.~(\ref{VusVcb}), the phase of $- V_{us} V_{cd} V_{tb}$ is of order $\mathcal{O}(\lambda^4) \sim 0.2 \%$.
At this point, it seems meaningful to examine the remaining two invariants; 
\begin{align}
\arg \left[ - \frac{  V_{ub} V_{cs} V_{td} }{ \det V_{\rm CKM} } \right]
= 92.40^{\circ}  \fallingdotseq  \a  \, , ~~~ 
\arg \left[ \frac{  V_{ub} V_{cd} V_{ts} }{ \det V_{\rm CKM} } \right]
= -64.06^{\circ}  \simeq - \g \, . 
\end{align}
That is, three of the six invariants are approximately identified with the angles $\alpha - \pi$, $-\beta$, and $-\gamma$, and the remaining three  represent small differences between the angles and invariants. 
Although such third-order invariants have already been discussed in the form fixed by $\det V = 1$ \cite{Kuo:2005pf}, this scheme provides an alternative perspective on the characteristic CP phases in the CKM matrix. 

%%%%%%%%%%%%%%
\section{Summary}
%%%%%%%%%%%%%%

In this letter,  using a rephasing invariant formula 
$\d =  \arg [ { V_{ud} V_{us} V_{c b} V_{tb} / V_{ub} \det V_{\rm CKM} }]$, 
we evaluate the CP phase $\delta$ in the CKM matrix $V_{\rm CKM}$ perturbatively for small quark mixing angles $s_{ij}^{u,d}$ with associated phases $\r_{ij}^{u,d}$.
 Consequently, we derived a relation 
$\d \simeq \arg [\Delta s_{12} \Delta s_{23} / ( \Delta s_{13} - s^u_{12} e^{-i \rho^u_{12}}  \Delta s_{23} )]$
with $\Delta s_{ij} \equiv s^d_{ij} e^{-i \rho^d_{ij}} - s^u_{ij} e^{-i \rho^u_{ij}}$, 
which represents the analytic behavior of the CKM phase. 
The uncertainty in the relation is of order $O(\lambda^{2}) \sim 4\%$, comparable to the current experimental precision. 
Comparisons with experimental data suggest that the hypothesis of some CP phases being maximal.

We also discussed relationships between the phase $\d$ and unitarity triangles. 
As a result, several relations between the angles $\alpha, \beta, \gamma$ and $\delta$ are identified through other invariants $V_{il} V_{jm} V_{kn} / \det V_{\rm CKM}$. 
These general perturbative relations broadly cover phenomenological calculations, 
 and therefore, the presented results have wide applicability for studies of CP violation in flavor physics and unified theories. 

%%%%%%%%%%%%%%
\section*{Acknowledgment}
%%%%%%%%%%%%%%

The study is partly supported by the MEXT Leading Initiative for Excellent Young Researchers Grant Number JP2023L0013.

%\bibliographystyle{bib/h-physrev50}
%\bibliography{bib/fourzero,bib/onezero,bib/refsym,bib/mutausym,bib/PSGUT,bib/StrongCP,bib/LR,bib/GCP,bib/U(2),bib/flaxion,bib/minimal-natural,bib/chiral, bib/T2HK,bib/CKM2MNS,bib/KMCPV,bib/Kuo}

\end{document}